\documentclass[useAMS,usenatbib]{mn2e}
\usepackage{epsfig,bm}
\usepackage{amsbsy,amssymb,amsmath}
\usepackage{hyperref}

\newcommand{\alt}{\lesssim}
\providecommand{\eprint}[1]{\href{http://arxiv.org/abs/#1}{#1}}
\providecommand{\adsurl}[1]{\href{#1}{ADS}}
\providecommand{\ISBN}[1]{\href{http://cosmologist.info/ISBN/#1}{ISBN: #1}}
\newcommand{\Planck}{{\sc Planck}}
\newcommand{\RECFAST}{{\sc recfast}}
\newcommand{\CAMB}{{\sc camb}}
\newcommand{\CMBFAST}{{\sc cmbfast}}
\newcommand{\CosmoMC}{{\sc CosmoMC}}
\newcommand{\vtheta}{\boldsymbol{\theta}}

\newcommand{\eV}{\rm{eV}}
\newcommand{\ud}{\rm{d}}
\newcommand{\Mpc}{\text{Mpc}}
\newcommand{\Hunit}{~\text{km}~\text{s}^{-1} \Mpc^{-1}}
\newcommand{\bea}{\begin{eqnarray}}
\newcommand{\eea}{\end{eqnarray}}
\def\zmax{z_{\rm max}}
\def\zmin{z_{\rm min}}

\renewcommand{\la}{\langle}
\newcommand{\ra}{\rangle}
\newcommand{\apj}{Ap.\ J.}
\newcommand{\mnras}{MNRAS}

\title[CMB and the Ionization History]
{The Cosmic Microwave Background and the Ionization History of the Universe}
\author[A.M.~Lewis, J.~Weller and R.A.~Battye]
{Antony~Lewis$^{1}$\thanks{Formerly at CITA, 60 St. George St, Toronto
  M5S 3H8, ON, Canada; contact details at \url{http://cosmologist.info}.},
  Jochen Weller$^{2}$ and Richard~Battye$^3$\\$^1$Institute of
Astronomy, Madingley Road, Cambridge CB3 0HA, UK.
  \\$^{2}$Department of Physics and Astronomy, University College
  London, Gower Street, London WC1E 6BT, UK. \\$^3$Jodrell Bank
  Observatory, University of Manchester, Macclesfield, Cheshire SK11 9DL, UK.
}
\pagerange{\pageref{firstpage}--\pageref{lastpage}}

\begin{document}
\maketitle

\label{firstpage}

\begin{abstract}
Details of how the primordial plasma recombined and how the universe
later reionized are currently somewhat uncertain. This uncertainty can
restrict the accuracy of cosmological parameter measurements from the
Cosmic Microwave Background (CMB). More positively, future CMB data can be used to constrain the
ionization history using observations. We first discuss how current
uncertainties in the recombination history impact parameter
constraints, and show how suitable parameterizations can be used to
obtain unbiased parameter estimates from future data. Some parameters
can be constrained robustly, however there is clear motivation to
model recombination more accurately with quantified errors.  We then
discuss constraints on the ionization fraction binned in redshift
during reionization. Perfect CMB polarization data could in principle
distinguish different histories that have the same optical depth. We
discuss how well the \Planck\ satellite may be able to constrain the
ionization history, and show the currently very weak constraints from
WMAP three-year data.
\end{abstract}
\section{Introduction}
We are entering the era of precision cosmology, with future high
precision cosmic microwave background (CMB) data in the offing. If the
physics governing the evolution of the photon distribution can be
modelled reliably, this data offers an almost unique opportunity to
measure a large number of cosmological parameters accurately and
distinguish models of the early universe. It is widely recognized that
small second (and higher) order effects can effect the CMB power
spectra at the several-percent level: these include the kinetic and
thermal Sunyaev-Zel'dovich (SZ) effects~\citep{Hu:2001bc} and CMB
lensing~\citep{Lewis:2006fu}. As well as the cosmological parameters
of most interest, there are however also other uncertain parameters
governing the background evolution of the universe that can be
of comparable importance.

Here we focus on the ionization history, parameterized as the
ionization fraction as a function of redshift. The broad details of
recombination are well
understood~\citep{Peebles:68,Peebles:93,Hu:1995fq}. Direct
recombination to the ground state is ineffective as it releases a high
energy photon that immediately ionizes another atom: it can only be
important if the photon is sufficiently cosmologically redshifted
before encountering another atom. The dominant mechanism is in fact
capture to an excited state followed by a two-photon transition to the
ground state. However there are many excited states in both hydrogen
and helium, giving many possible recombination channels. Furthermore
level populations may be out of equilibrium, requiring a full
multi-level atom evolution to calculate the recombination rate
accurately~\citep{Seager:1999km}. This can only be done to the extent
that the different transition rates are known (or can be calculated),
in particular excited state two-photon transition rates are not known
very well and are potentially important~\citep{Dubrovich:2005fc}. The
two-photon rates can also differ significantly from their empty-space
value due to induced decay by CMB photons~\citep{Chluba:2005uz}.

Most current CMB power spectrum calculations use the effective model
used in the code \RECFAST~\citep{Seager:1999bc}. However the
ionization fraction from this code differs by several percent from a
more recent calculation in \citet{Dubrovich:2005fc} that includes
additional transitions. Level splitting and other neglected effects can also have
percent-level effects on the CMB power
spectra~\citep{Chluba:2005uz,Leung:2003je,Rubino-Martin:2006ug}. Furthermore there is no
full calculation with any quantification of the error. When high
resolution CMB data are available we will either need to be able to
calculate the recombination history sufficiently accurately that
errors can be neglected, or we will need to include uncertainties in
any analysis. Not surprisingly assuming an incorrect history can give
biased constraints on the cosmological parameters inferred from CMB
data~\citep{Hu:1995fq}. Our aim in this paper is to see whether a
crude parameterization of uncertainties in recombination can be used
with future data to give reliable parameter constraints when
considerable uncertainties remain in the details of recombination. The
detailed task of improving recombination models and parameterizing
residual uncertainties is an important challenge for the future.

Once recombination has proceeded to reduce the residual ionized
density to a low level the details are no longer important for the
CMB: the electron density is so low that few CMB photons are
scattered. However at some point first collapsed objects will form,
and at some later time high-energy photons emitted by (for example)
quasars or stars of various populations can cause the universe to reionize
\citep{Loeb:2000fc,Gnedin:2006uz}. Current observations of quasar absorption spectra
only put a lower bound on reionization redshift, giving evidence for
the first neutral hydrogen on our light cone at $z\approx 6$
\citep{Becker:2001ee,Fan:2006dp}. Exactly how the ionization fraction
evolved between the low level remaining after recombination and
$z\approx 6$ is unknown in any detail.

The main CMB constraint on reionization comes from the large scale
polarization signal generated by scattering of the CMB quadrupole
during reionization. This gives a characteristic bump in the large
scale polarization power spectra on scales larger than the horizon
size at reionization, and the exact details of the shape of the bump
can in principle be used to constrain the ionization history
\citep{Kaplinghat:2002vt,Hu:2003gh}. Scales smaller than the horizon
size are uniformly damped, leading to a suppression of the acoustic
peaks of $e^{-2\tau}$ where $\tau$ is the optical depth to
reionization; the small scales cannot therefore be used to constrain
the details of the history, only the total optical depth. Beyond
linear theory there are also tiny secondary anisotropies on  small
scales that we do not discuss further here
\citep{Weller:1999ch,Hu:1999vq}.

The current constraint on the optical depth from the WMAP three-year
polarization observations is $\tau = 0.09\pm
0.03$~\citep{Page:2006hz,Spergel:2006hy}, significantly smaller than
$\tau\sim 0.17$ favoured by the one-year data (which had unsubtracted
foreground contamination). The new value may be more consistent with
simple reionization scenarios, though the lower fluctuation amplitude
measured by the three-year data makes this interpretation unclear; for
discussion see \citet{Alvarez:2006ma,Popa:2006rh}.

Since there is currently no convincing model of the reionization
history we attempt to constrain it in a
relatively model independent way. \citet{Hu:2003gh} suggest a binned
fit and performed a principal component analysis with Fisher matrix
techniques. They showed that the first two to three
eigenmodes should be fairly well constrained with future observations,
but other details are effectively unconstrained by the CMB alone. In
the second part of this paper we investigate  how the reionization
history binned in redshift can be constrained with current, perfect,
and simulated \Planck~\citep{unknown:2006uk} data. The general
constraints we obtain are rather weak, however distinct models of
reionization can be distinguished. Unless there is a convincing
physical model for how reionization proceeds, flexible modelling of
the reionization history is required in order not to obtain biased
constraints on the optical depth (and hence the amplitude of
primordial fluctuations and $\sigma_8$ inferred from the CMB).

Since details of reionization are only important on large scales, and
details of recombination are only important on small scales, the two
effects are virtually independent. We start by describing our
parameter estimation and forecasting methodology, then move on to
consider recombination and reionization separately.

\section{Parameter estimation and forecasting}
\label{sec:MCMC}
\newcommand{\mCh}{\hat{\bm{C}}}
\newcommand{\mC}{\bm{C}}
\newcommand{\begm}{\begin{pmatrix}}
\newcommand{\enm}{\end{pmatrix}}

We make use of standard Markov Chain Monte Carlo (MCMC) methods to
sample from the posterior distribution of the parameters given real or
forecast data. Extra parameters governing the ionization history are
added to the code \CAMB~\citep{Lewis:1999bs} (based on
\CMBFAST~\citep{Seljak:1996is}) for computing the CMB
anisotropies. This is then used with a modified version of
\CosmoMC~\citep{Lewis:2002ah} to generate samples from the posterior
distribution.

For idealized forecasting work we assume that the temperature $T$ and polarization $E$ fields
(including noise) are
statistically isotropic and
Gaussian, and take the $B$-polarization signal to be negligible. We
shall assume that foreground sources of polarization can be accurately
subtracted using multi-frequency observations.
The covariance over realizations is given by
\begin{eqnarray}
\mC_l &\equiv& \left\la \begin{pmatrix} T_{lm} \\ E_{lm} \end{pmatrix}
  \begin{pmatrix} T_{lm}^* &  E_{lm}^* \end{pmatrix} \right\ra \\
  \nonumber &=&
\begin{pmatrix} C^{TT}_l + N^{TT}_l& C^{TE}_l \\ C^{TE}_l & C^{EE}_l +
N^{EE}_l\end{pmatrix},
\end{eqnarray}
where $N_l$ is an assumed isotropic noise contribution, and we take the
noise on $E$
and $T$ to be uncorrelated.

For a given sky realization one can construct estimators of the $C_l$
given by
\begin{equation}
\hat{C}_l^{XY} \equiv \frac{1}{2l+1} \sum_{m} X_{lm}^* Y_{lm}
\end{equation}
so that $\la \hat{C}_l^{XY} \ra = C_l^{XY} + N^{XY}_l$.
The likelihood for the matrix of the estimators $\mCh_l$ given a
theoretical matrix $\mC_l$, assuming
$T_{lm}$ and $E_{lm}$ are Gaussian, is then
\begin{equation}
-2\log P(\mCh_l|\mC_l) = (2l+1) \left\{\text{Tr}\left[ \mCh_l \mC_l^{-1}
  \right] + \log |\mC_l| \right\}.
\end{equation}
We calculate the expected log likelihood for each set of parameters
$\vtheta$. For some fiducial model with parameters $\vtheta_0$ this is
given by $\la \log P(\vtheta | \text{data})\ra$ where the average is
over data realizations that could come from the $\vtheta_0$
model. Hence the distribution we sample from is the exponential of
\begin{equation}
\la \log P(\vtheta|\vtheta_0) \ra =-\frac{1}{2}\left(
\text{Tr}\left[\mC(\vtheta_0)\, \mC^{-1}(\vtheta)\right] + \log
|\mC(\vtheta)|\right),
\end{equation}
where the covariance matrix now includes all the different $l$, $m$
modes (we use $2\le l \le 2000$).
The mean log likelihood peaks at the true model $\vtheta=\vtheta_0$, and
the shape of the likelihood encapsulates any important degeneracies in
the data. To this extent our method is superior to Fisher-based
estimates that can be misleading if the posterior is significantly
non-Gaussian. Posteriors obtained using the mean log likelihood method
are
approximately the same width as those obtained in most actual realizations
of $\mCh$ close to the fiducial model.
Our method of using MCMC has the benefit of being immediately
applicable to real data, allowing us to test much of the parameter
estimation pipeline for consistency by forecasting. Note that even
though the mean log likelihood peaks at the true model, if the
posterior is non-Gaussian  marginalized constraints on individual
parameters need not peak at the true model values.

For our fiducial models we take the best fit six parameter WMAP
three-year concordance $\Lambda$CDM model \citep{Spergel:2006hy}. The
parameters we vary are the baryon density $\Omega_b h^2$, dark matter
density $\Omega_c h^2$, approximate ratio of the sound horizon to the
angular diameter distance at last scattering $100\theta$ (from
which we derive the Hubble parameter $H_0=100h\Hunit$), constant
scalar adiabatic spectral index $n_s$ and scalar adiabatic amplitude
$A_s$ at $k=0.05\Mpc^{-1}$ parameterized with a flat prior on
$\log(10^{10}A_s)$. When considering recombination we assume sharp
reionization with
optical depth $\tau$ (fiducial value $\tau=0.091$). We neglect the
neutrino masses, and assume negligible tensor and non-adiabatic
modes. Fiducial values for other model parameters are $\Omega_b
h^2=0.0223$, $\Omega_c h^2 = 0.104$, $H_0=73$, $n_s=0.955$,
$\log(10^{10} A_s)=3.02$.
We assume a fixed helium fraction of $0.24$; marginalizing over
uncertainties in this parameter would be trivial, with the effect depending on what (if any) external constraint is assumed.

For the \Planck~satellite we take a toy model with isotropic noise $N^{TT}_l = N^{EE}_l/4 = 2~~
\times
10^{-4} \mu K^2$ on large scales, with an effective Gaussian beam width of
$7$ arcminutes \citep{unknown:2006uk}. Since our purpose here is to
isolate the effects of reionization we shall not include complicating
secondary signals such as SZ or CMB lensing, though these must of
course be accounted for when analysing real future data. The lensing
effect can be included easily enough and, once included consistently,
has little effect on the recovered parameters at \Planck\ sensitivity
\citep{Lewis:2005tp}. Thermal SZ can in principle be removed using the
frequency information. Kinetic SZ is more difficult to
model~\citep{Zahn:2005fn}, though expected to be a small signal at
$l\alt 2000$; future work is required to model it sufficiently
accurately for reliable parameter estimation. In principle it might be
necessary to model the SZ signal as a function of reionization
parameters since the kinetic SZ signal comes from the inhomogeneous
reionization epoch. Indeed ultimately the SZ signal may provide a
useful constraint on the ionization history, though here we focus on
what can be learnt using only the linear polarization signal.

\section{Recombination}
\begin{figure}
\psfig{file=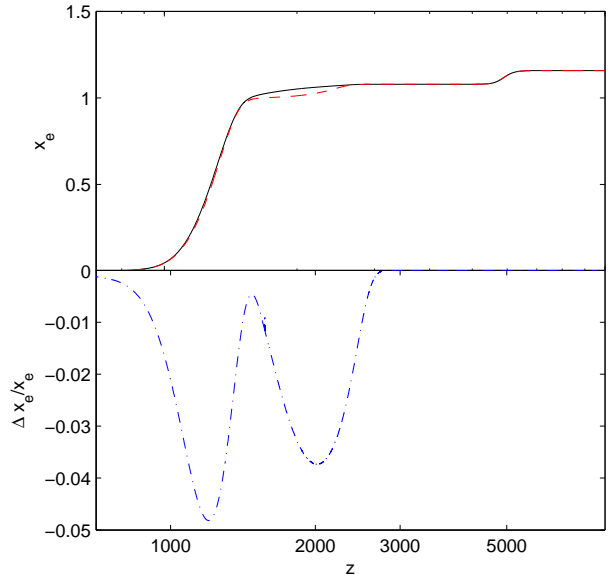,width=8cm}
\caption{The ionization fraction as a function of redshift during
  recombination. The solid line is computed using \RECFAST\
  \citep{Seager:1999km}, the dashed line includes a model of
  additional transitions from~\citet{Dubrovich:2005fc}. The bottom
  panel shows the several-percent fractional difference.}
 \label{fig:recomb}
\end{figure}

\subsection{Uncertainties in the standard recombination}
Before discussing the effect of uncertainties in the recombination
history, we will briefly review the the standard recombination
dynamics as implemented in the widely used \RECFAST\ code. For full
accuracy one has to do a radiative transfer calculation accounting for many different levels of
the hydrogen and helium atoms. However, for simplicity, we follow the
notation of \RECFAST\ and describe the recombination process for an
{\em effective 3-level} atom, i.e. a two-level atom
plus continuum. Detailed balance equations for
the proton fraction $x_{\rm p} =
n_{\rm p}/n_{\rm H}$ and singly ionized helium fraction $x_{\rm HeII} =
n_{\rm HeII}/n_{\rm H}$ lead to~\citep{Seager:1999bc}:
\begin{eqnarray}
\label{eq:newstandard_xe}
{dx_{\rm p}\over dz} &= \left(x_{\rm e}x_{\rm p} n_{\rm H} \alpha_{\rm H}
 - \beta_{\rm H} (1-x_{\rm p})
   {\rm e}^{-h\nu_{H2s}/kT_{\rm M}}\right) \\
 &\times\quad{\left(1 + K_{\rm H} \Lambda_{\rm H} n_{\rm H}(1-x_{\rm p})\right)
    \over H(z)(1+z)\left(1+K_{\rm H} (\Lambda_{\rm H} + \beta_{\rm H})
     n_{\rm H} (1-x_{\rm p}) \right)},\nonumber
\end{eqnarray}
and
\begin{equation}
\label{eq:HeI_xe}
\begin{array}{l}
\displaystyle
{dx_{\rm HeII}\over dz} =   \frac{1}{H(z)(1+z)}
\\
 \\
\displaystyle
   \times \left(x_{\rm HeII}x_{\rm e} n_{\rm H} \alpha_{\rm HeI}
   - \beta_{\rm HeI} (f_{\rm He}-x_{\rm HeII})
   {\rm e}^{-h\nu_{HeI2^1s}/kT_{\rm M}}\right) \\
\\
\displaystyle
 \times {\left(1 + K_{\rm HeI} \Lambda_{\rm He} n_{\rm H}
  (f_{\rm He}-x_{\rm HeII}){\rm e}^{-h\nu_{ps}/kT_{\rm M}})\right)
  \over\left(1+K_{\rm HeI}
  (\Lambda_{\rm He} + \beta_{\rm HeI}) n_{\rm H} (f_{\rm He}-x_{\rm HeII})
  {\rm e}^{-h\nu_{ps}/kT_{\rm M}}\right)},
\end{array}
\end{equation}
where $x_e \equiv n_e/n_{\rm H}$ is the ionization fraction, $n_{\rm H}$
is the total density of ionized and neutral hydrogen. The relation
between the recombination coefficients $\alpha$ and the
photoionization coefficients $\beta$ is given by $\beta=\alpha (2\pi m_{\rm e} k \
T_{\rm M}/h^2)^{3/2}
\exp(-h\nu_{2s}/kT_{\rm M})$. The
frequencies $\nu$ are from the characteristic wavelength of the atomic
transitions under consideration as indicated by their indices. Note
that the hydrogen recombination rate $\alpha_{\rm H}$ in \RECFAST\ includes a fudge parameter $F$ to effectively describe the multilevel
atom. $f_{\rm
  He}$ is the He/H number ratio. The
terms in brackets of expression \eqref{eq:newstandard_xe} and
\eqref{eq:HeI_xe} are from the detailed balance between recombination
and photoionization, while the multiplied terms take into account
redshifting of the Lyman-$\alpha$ photons for the hydrogen atom and
for HeI the $2^1p-1^1s$ photons via the $K$-factors. Further the
rates of the two photon decays are given by $\Lambda_{\rm H}$ and
$\Lambda_{\rm He}$ respectively. The recombination and photoionization
rates depend on the temperature of the baryons and photons, where the
evolution of the baryon temperature $T_{\rm M}$ is given by
\begin{equation}
\label{eq:cooling}
 \frac{dT_{\rm M}}{dz} = \frac{8\sigma_{\rm T}a_{\rm R}
   T_{\rm R}^4}{3H(z)(1+z)m_{\rm e}c}\,
  \frac{x_{\rm e}}{1+f_{\rm He}+x_{\rm e}}\,(T_{\rm M} - T_{\rm R})
  + \frac{2T_{\rm M}}{(1+z)},
\end{equation}
where $T_{\rm R}$ is the radiation temperature, $a_{\rm R}$ is the  radiation constant,  and the Thomson scattering cross section
$\sigma_{\rm T}$. For further details and the exact values of the
constants see~\citet{Seager:1999km,Seager:1999bc}.

We can now discuss two slightly different calculations of the
recombination history as shown in Fig.~\ref{fig:recomb}. One
calculation uses \RECFAST, the other is altered at the
several-percent level by allowing for additional transitions. This is
achieved using the model of \cite{Dubrovich:2005fc} by
modifying the two-photon Einstein coefficients
$\Lambda$ to allow transitions from upper levels and making them
temperature dependent, i.e.
\begin{equation}
\Lambda \to
\Lambda+\sum_{i=i_0}^{i_N}g_iA_{i1}^{(2q)}e^{h(\nu_{iC}-\nu_{2s,c})/kT_{M}}\; ,
\end{equation}
where the sum runs over the upper levels, $\nu$ are the particular
transition frequencies and $A_{i1}^{(2q)}$ the upper level Einstein
coefficients with $g_i$ the statistical weights of the states. The $K$ factors also have a less important modification following \cite{Dubrovich:2005fc}.
This particular modification may not be at all accurate, but we can take the
difference between these two results to indicate the level of current
theoretical uncertainty in the recombination history. Accounting for
more transitions generally increases the recombination paths and hence
speeds up recombination. One of the main effects is to slightly shift
the redshift of maximum visibility, and hence the angular scale of the
CMB acoustic peaks. However the exact shapes of the CMB power spectra
are sensitive to the full shape of the recombination curve: the small
scale peak suppression and polarization are quite sensitive to the
width of recombination. For percent-level accuracy in the power
spectra, the ionization fraction needs to be known to percent-level or
better through the peak of the visibility.

\begin{figure}
\psfig{file=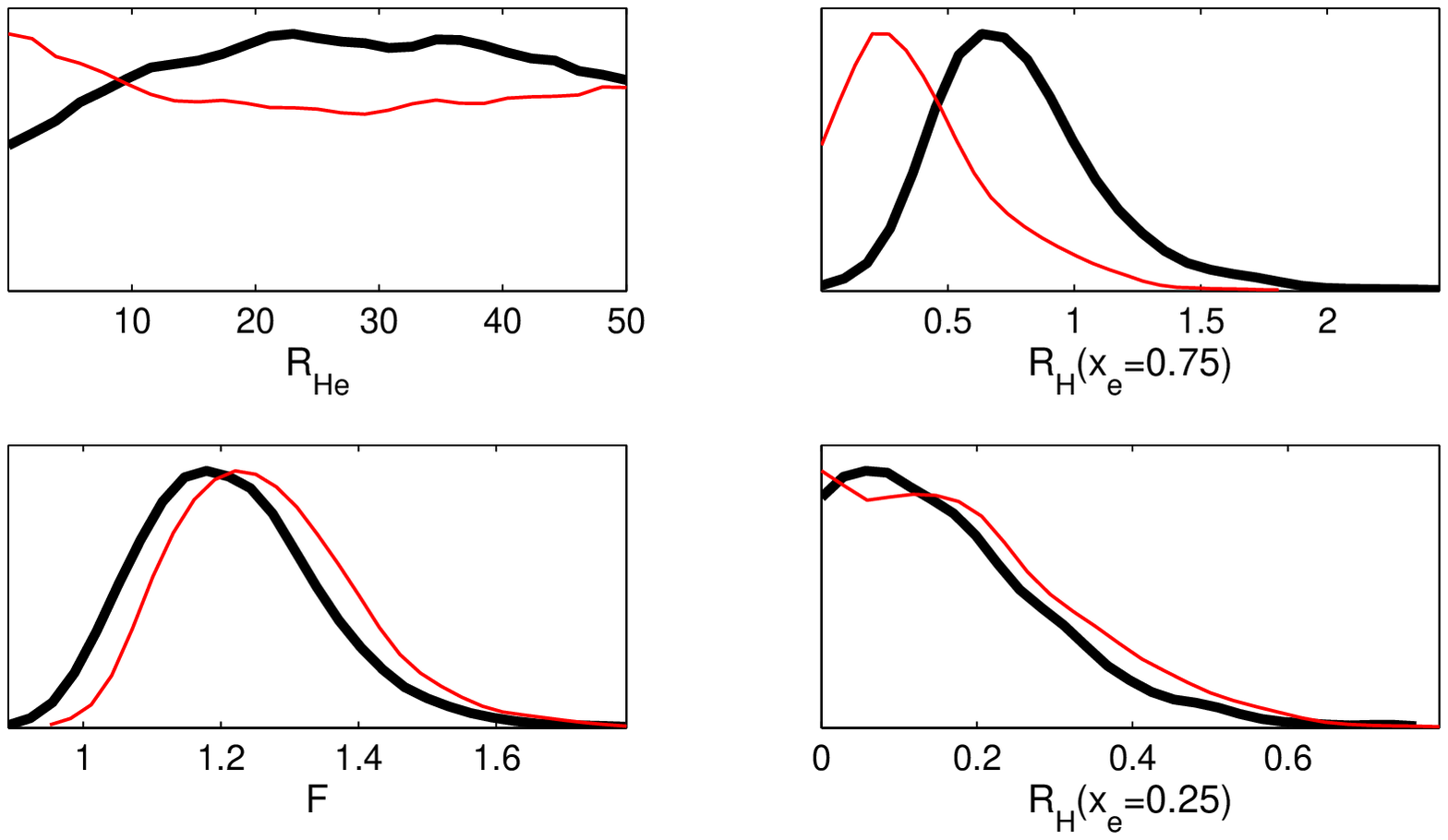,width=8cm}
\psfig{file=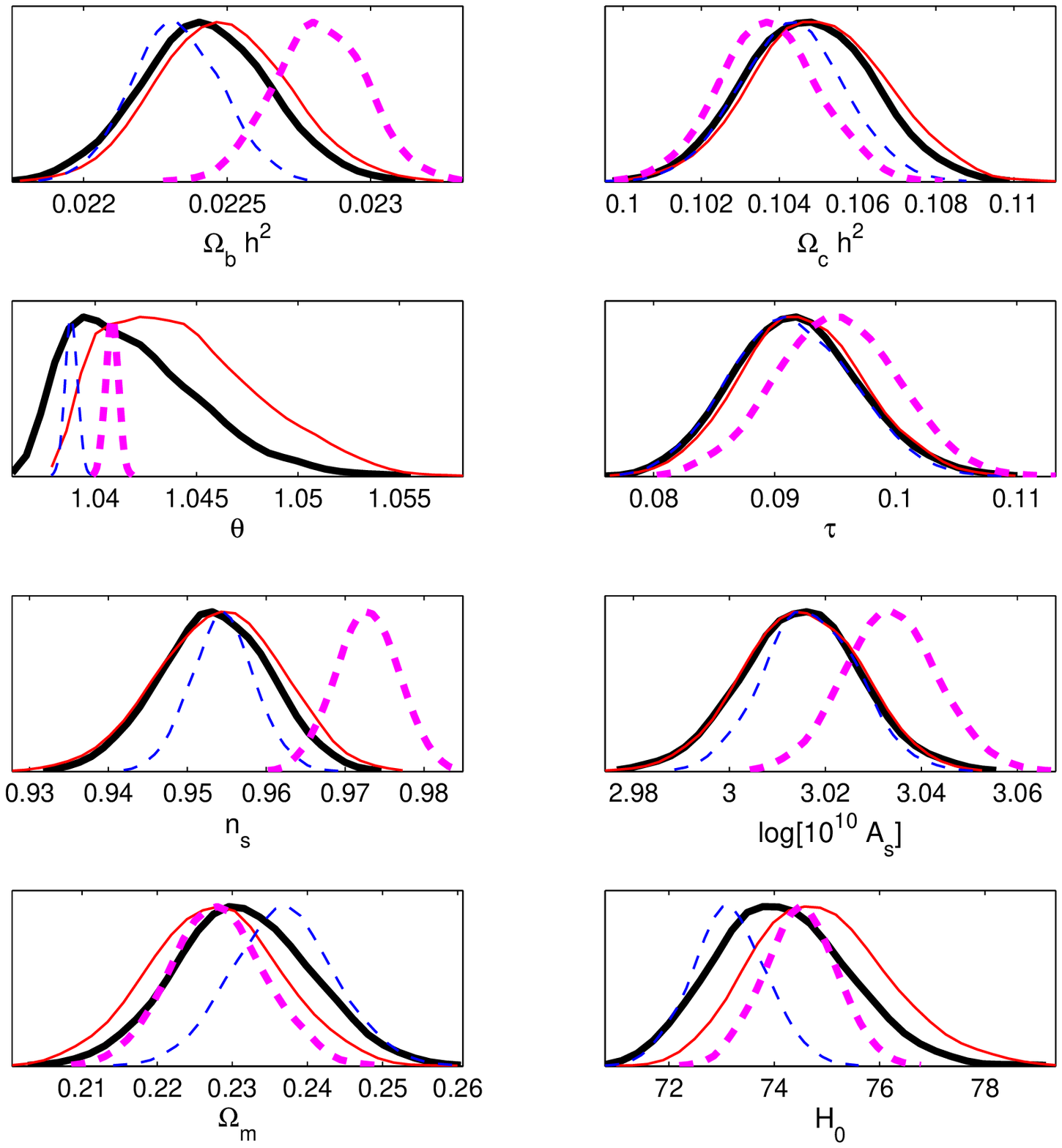,width=8cm}
\caption{Forecast \Planck\ constraints on recombination and cosmological parameters.
There are two fiducial models, one using \RECFAST\ (thin lines) and one using \RECFAST\ with extra transitions following \citet{Dubrovich:2005fc} (thick lines). Each simulated data set is analysed in two ways:  dashed curves
show the constraints using \RECFAST\ and no additional recombination parameters; solid lines show the constraints allowing for four extra effective recombination parameters, as described in the text. The dotted lines show that using a fixed wrong recombination model can give biased parameter constraints; allowing for extra parameters (solid lines) broadens the errors bars but gives consistent results for both fiducial models.}
\label{fig:recomb_planck}
\end{figure}

Helium recombination is more complicated than hydrogen, but also less
important: the significant difference between the HeI recombination
histories only has a percent-level effect on the observed CMB spectrum
on small scales due to the slightly modified diffusion damping
length.

The dashed curves in Fig.~\ref{fig:recomb_planck} show the bias on
parameter constraints from \Planck\ if an incorrect recombination
history is used that differs by as much as the
\citet{Dubrovich:2005fc} corrections to \RECFAST. It should therefore
be a matter of some priority to continue the work of
~\citet{Seager:1999km} and \citet{Dubrovich:2005fc,Chluba:2005uz,Rubino-Martin:2006ug} to
include all transitions that might be important, and to quantify the
importance of poorly known transition rates on the predictions.

In this paper we investigate the use of a crude ad hoc
parameterization that can be used to obtain correct cosmological
parameter constraints given the current level of uncertainties. We
regard this as a proof of principle rather than a recommendation for
practice. It may be a useful guide to where most effort needs to be
concentrated to obtain more accurate results in future.

We base our calculations on the \RECFAST\ code.
We evolve these same equations, with modifications incorporating
additional parameters. In \RECFAST\ the effect of out of equilibrium
distributions lead to a faster rate of hydrogen recombination than in
a simple analysis. This is accounted for by multiplying the
recombination coefficient by a `fudge factor' F, fixed to 1.14 in
\RECFAST\ to match their full multi-level atom results. Other authors get slightly different values~\citep{Rubino-Martin:2006ug}. We take F to
be a free parameter that effectively governs the speed of the end of
recombination.

At earlier times recombination is dominated by the
two-photon transitions, accounted for in \RECFAST\ only from the
lowest excited state. The calculation of \citet{Dubrovich:2005fc}
includes additional transitions from higher states, and there are also
corrections from induced decay~\citep{Chluba:2005uz}. We parameterize
the uncertainties governing the two-photon transitions by using an
effective two-photon rate $\Lambda_{\rm H}$ parameterized by
\begin{equation}
\Lambda_{\rm H} = \Lambda_{\rm H}^{(2)}\left[ 1 + R_H(x_e=0.75) \left(\frac{R_H(x_e=0.25)}{R_H(x_e=0.75)}\right)^{2(0.75-x_e)}\right].
\end{equation}
Here $\Lambda_{\rm H}^{(2)}$ is the (fixed) empty-space 2s-1s two-photon
rate, and the two parameters $R_H(x_e=0.75)$ and $R_H(x_e=0.25)$
measure an additional early and late contribution to the two-photon
rate from extra transitions and induced decay.
This model is not well motivated physically, but has the benefit of
being fairly well constrained from the data. The exponential form
allows it to capture exponential behaviour from Boltzmann-factors
governing populations of excited states.
Since the CMB is only weakly sensitive to the Helium recombination, we
use the single parameter $R_{\text{He}}$ so that the effective
two-photon Helium transition rate is given by $\Lambda_{\text{He}}
=\Lambda_{\text{He}}^{(2)}R_{\text{He}}$. We do not change anything
else. Although rather ad hoc, we expect that these parameters should
be able to describe most changes to the recombination rate, and have a
physical interpretation as measuring an effective rate at four
different levels of ionization. Alternative parameterizations are
discussed by \citet{Hannestad:2001gn}.

Fig.~\ref{fig:recomb_planck} shows the parameter constraints from our
\Planck\ mean log likelihood forecasting function, using two different fiducial models of recombination. Including
marginalization over the four additional recombination parameters
allows the cosmological parameters to be extracted correctly in both
cases, though with increased error bars. The posterior of the
recombination parameters clearly shows that $R_H(x_e=0.75)$ and $F$
can be constrained away from zero, though the other two parameters are
poorly constrained. With good polarization data the parameter $F$ is
in fact very benign: varying only $F$ recovers cosmological parameters
with essentially the same error bars as fixing it to 1.14 (the
\RECFAST\ value). However varying the rate of the two-photon hydrogen
transitions shifts the redshift of maximum visibility, hence the large
increase in the posterior range of $\theta$ that governs the
positions of the acoustic peaks. Since our parameterization only
allows for speeding up of transitions relative to \RECFAST, the
posterior distributions are non-Gaussian with a tail to larger
$\theta$  corresponding to faster recombination, and a one-sided
increase to the matter density posterior.

We have only considered basic 6-parameter cosmological models here,
and more general models may be more degenerate with the recombination
parameters, though a running spectral index can be constrained well
even with the extra parameters.

Since details of recombination only significantly affect the small
scale power spectrum, the impact of standard recombination
uncertainties is rather mild when considering only the current WMAP
three-year data: parameter constraints are affected only by a fraction
of the error bar, comparable to the effects of different priors,
likelihood modelling approximations, and secondary signals. For this
reason we have not presented constraints on the recombination
parameter space for WMAP; \RECFAST\ is sufficiently accurate to obtain
reliable parameters at the moment.

\subsection{Non-standard models}

\begin{figure}
\begin{center}
\epsfig{file=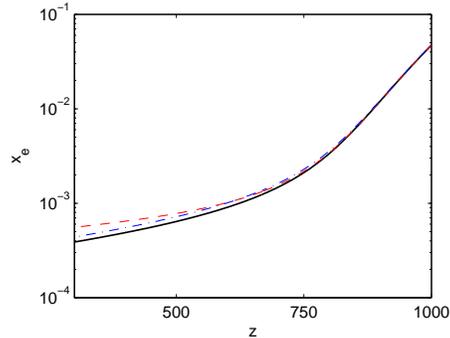,width=6cm}
\caption{Ionization history for standard \RECFAST\ (fudge parameter
  $F=1.14$, solid), \RECFAST\ with $F=1$ (dash-dotted), and \RECFAST\
  with $F=1.14$ and an input of $3\times 10^{-24}\eV s^{-1}$ per
  proton today from homogeneous dark matter annihilation (dashed; see
  \citet{Padmanabhan:2005es}, equivalent to $F_{26}=0.06$ of
  \citet{Zhang:2006fr}).}
\label{fig:nonstandard}
\end{center}
\end{figure}

\begin{figure}
\begin{center}
\epsfig{file=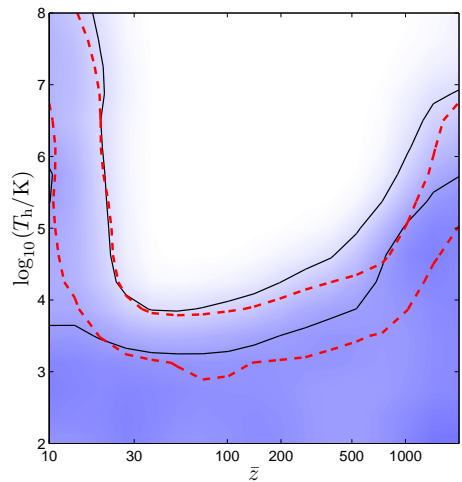,width=6cm}
\caption{WMAP (solid line) and simulated PLANCK (dashed line)
  constraints on the parameter $T_{\rm h}$ as a function of redshift
  $\bar z$. Shading corresponds to the marginalized probability for
  the WMAP constraint and contours are at $68\%$ and $95\%$.}
\label{fig:heat}
\end{center}
\end{figure}

Our ad hoc parameterization may also pick up (and partly correct for)
many non-standard processes, for example annihilating dark matter
could inject energy during and after recombination.  Low values for
the free fudge parameter $F$ would be likely to fit these models
better, corresponding to the increased ionization fraction at the end
of recombination in these scenarios
\citep{Chen:2003gz,Padmanabhan:2005es,Zhang:2006fr}.
Figure~\ref{fig:nonstandard} shows an ionization history with dark
matter annihilation  from \citet{Zhang:2006fr}, compared to a
\RECFAST\ history with lower value of the fudge parameter $F$. The
shape at the tail end of recombination is somewhat similar, even though the
(very small) ionization fraction at late times is significantly
different. The resulting $C_l$ agree to below a percent except for
few-percent change to the large-scale polarization signal.
As we have seen $F$ can be constrained well, largely as a result of
\Planck's good polarization sensitivity, so models with significant
annihilation should also be clearly distinguishable. If we treat the
annihilation rate as a free parameter, but fix recombination otherwise
to \RECFAST, our forecast for \Planck\ suggests homogeneous rates
today of $\gtrsim 3\times 10^{-24}\eV s^{-1}$ per proton will be
detectable at $95\%$-confidence, consistent with the forecast of
\citet{Padmanabhan:2005es}. This is also consistent with the
constraint on the fudge parameter $F$ shown in
Figure~\ref{fig:recomb_planck} when comparing the annihilation history
with the similar $F=1$ model shown in Figure~\ref{fig:nonstandard}.

 If the annihilation rate is much higher than that shown in
 Fig.~\ref{fig:nonstandard} the fudge parameter $F$ will no longer be
 able to mimic the decay well. This is because higher annihilation
 rates give a significant contribution to the optical depth from the
 lower redshift ionization fraction, and also change the $l\lesssim
 100$ polarization signal more radically. Such large annihilation
 rates should however be clearly distinguishable using an appropriate
 model.

Physically motivated dark matter candidates are generally expected
have a small effect on the CMB power spectra due to decay or
annihilation \citep{Mapelli:2006ej}. Constraints on dark matter
annihilation from recent data via the effect on the ionization history
are given in \citet{Zhang:2006fr,Mapelli:2006ej}, so we do not
consider non-standard models any further here.

Ad hoc models for the injection of Lyman-$\alpha$ photons during
recombination could also delay the time of maximum visibility
\citep{Peebles:2000pn,Doroshkevich:2002wy,Bean:2003kd}, so more
general parameterizations allowing for slower recombination would be
needed to pick up and account for more general models. Models which
increase the ionization fraction significantly at quite low redshift
can give rise to significant optical depth
\citep{Peebles:2000pn,Doroshkevich:2002wy,Bean:2003kd}. These models
are likely to be ultimately quite well constrained by the large scale
polarization signal as part of the recombination component.

An alternative way of modifying the precise details of recombination
is to increase the temperature of the matter (baryons), $T_{\rm M}$,
by the injection of energy, assuming the production of Lyman-$\alpha$
photons to be inefficient. Such a possibility has been considered
previously in \citet{Weller:1998ah} and achieves modifications to the
ionization history by shifting the balance between recombination and
cooling. A simple phenomenological model would be to consider the
input of energy with a Gaussian profile centred around $\bar z$ with
width $\rho$. This can be achieved by inclusion of an extra term in
\RECFAST\ (Eq.~\eqref{eq:cooling})
\begin{equation}
{dT_{\rm M}\over dz}\bigg|_{\rm h}={T_{\rm h}\over\rho}\sqrt{2\over\pi}\exp\bigg[-{2(z-\bar z)^2\over \rho^2}\bigg]\,,
\end{equation}
which, in the absence of any cooling processes and assuming that
${\bar z}>>\rho$, would lead to an increase in the matter temperature
$\Delta T_{\rm M}=T_{\rm h}$. This would require the input of a total
energy of $8.6{\rm eV}(T_{\rm h}/10^{5}{\rm K})$ per baryon over
the period of heating. In any given scenario cooling processes will
suppress the effect substantially, but it is still possible to
increase the temperature of the IGM enough for it to have an
observable effect on the CMB anisotropies.

We have investigated the constraint imposed by the current WMAP data
on $T_{\rm h}$ as a function of ${\bar z}$, marginalizing over
$0.05<\rho/{\bar z}<0.3$ and the results are presented in
Fig.~\ref{fig:heat}. We see that for late (${\bar z}<30$)  and early
times (${\bar z}>1100$) large amounts of energy ($T_{\rm h}>10^{7}{\rm
  K}$)  can be input without changing the CMB anisotropies; at low
redshift heating effects are swamped by  reionization. Large heat
inputs may be constrained in other ways, for example, by spectral
distortions in the CMB, however any reasonable heat input is not
constrained by observational limits on spectral distortions
\citep{Weller:1998ah}. There are tight constraints of $T_{\rm
  h}<10^{4}{\rm K}$ for $30 < \bar z <100$, $T_{\rm h}<10^{5}{\rm K}$
for $100<{\bar z}<500$ and $T_{\rm h}<10^{6}{\rm K}$ for $z\sim 1000$
(all limits are $2\sigma$).

By considering a fiducial model with $T_{\rm h}=0$, we have also used
simulated \Planck\ likelihood functions to show that more stringent
constraints will be possible in the near future in the range
$200<{\bar z}<1500$, as show by the dashed lines in Fig.~\ref{fig:heat}.

\section{Reionization}
\subsection{Binning the ionization fraction}
\begin{figure}
\epsfig{file=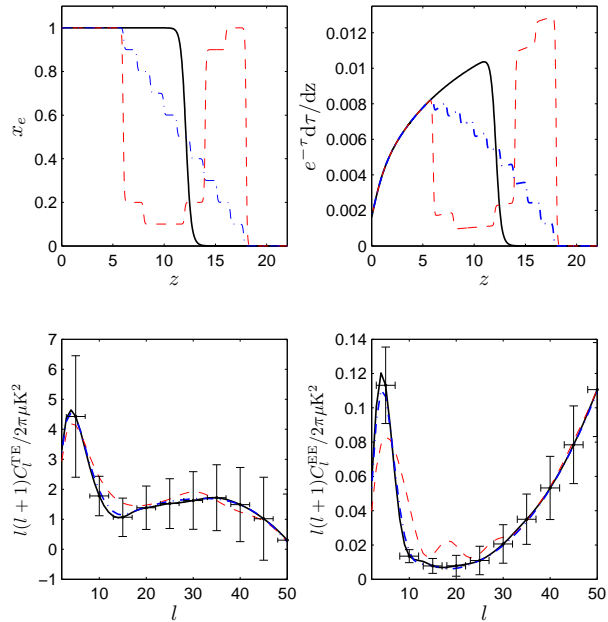,width=8cm}
\caption{CMB anisotropies and ionization histories for the fiducial
models used in this analysis: sharp (solid), dashed (double
reionization) and dragged (dash-dotted). The total optical
depth for all three models is $\tau = 0.1$. The top left panel is the
ionization fraction $x_e$, top right is the visibility (with respect
to $z$).
Bottom left are the temperature-polarization cross-correlation power
spectra, bottom right the $E$-polarization power spectra. Error bars
on the bottom plots show the noise plus cosmic variance  from our
simple model of \Planck\ in the sharp reionization model.}
\label{fig:fiducialmodels}
\end{figure}

In order to constrain the reionization history of the Universe we bin
the ionization fraction in redshift bins with
\begin{equation}
x_e(z) = x_i\; , \qquad  z_i-\frac{\Delta z}{2} < z < z_i+\frac{\Delta
  z}{2}\; .
\end{equation}
We further introduce a maximum redshift $\zmax$ above we follow the
standard recombination history without reionization and a minimum
redshift $\zmin$ below which we assume complete reionization. In
practice we require the ionization history to be smooth, so we join
the bins using a $\tanh$-function. We neglect Helium reionization at
$z\sim 3$ as this only has a very small effect on the CMB. Note that
since a constant $x_e$ electron density falls off as $n_e\propto
(1+z)^3$ in matter domination, the visibility per redshift interval
scales like $(1+z)^{1/2}$, and hence $x_e$ bins of fixed redshift
width contribute somewhat more to the optical depth at higher
redshift. The choice of best bin widths and offset is not obvious
without any clear theoretical priors on the expected reionization
history. These could be taken as additional parameters, but for
simplicity we fix them here; we choose the first bin to start at
$z\sim 6$ where we think the ionization fraction first was last
significantly less than unity. If the binning is in phase with any
features in the ionization history the reconstruction will be much
cleaner. It may therefore be a good idea to use two separate
reconstructions offset by half a bin for comparison.

We consider three fiducial reionization histories that all have an optical depth $\tau \approx 0.1$. The first is a ``standard'' reionization history with sharp complete reionization at a redshift $z\sim 12$. The second scenario is a dragged out reionization and the third a double reionization scenario (for possible physical models and discussion see \citet{Cen:2002zc,Furlanetto:2004nt}). The latter two are modelled using a series of narrow bins in redshift adjusted to give the same total optical depth. Other cosmological parameters are set to the values for the best fit six parameter WMAP three-year concordance model
\citep{Spergel:2006hy}.

\begin{figure}
\epsfig{file=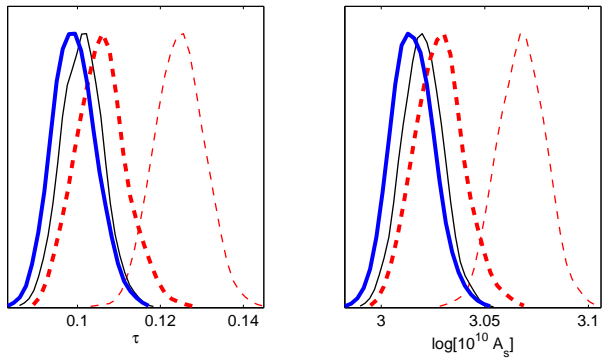,width=8cm}
\caption{\Planck\ optical depth and amplitude constraints from the
  sharp model analysed using the sharp model (thin solid), the
  incorrect result from analysing a double reionization model using a
  sharp model (thin dashed), and the consistent result from the double
  reionization (thick dashed) and sharp (thick solid) models using a
  binned reconstruction.}
\label{fig:reion_wrongparams}
\end{figure}

Figure (\ref{fig:fiducialmodels}) shows the reionization histories
and polarization power spectra for the three fiducial models. The
temperature power spectra are virtually identical, so polarization
information is essential. The accuracy with which the spectra can be
measured is limited by cosmic variance on these large scales even if
noise and foreground uncertainties were negligible. This limits the
amount of information that can be extracted from the CMB even with
perfect data. The fast and dragged reionization histories give rather
similar power spectra, however the double reionization history, which
has two distinct peaks in the visibility function, gives a clearly
different prediction (for further discussion see
\citet{Hu:2003gh}). If fast reionization is assumed, the inferred
optical depth would be incorrect if in reality there is double
reionization~\citep{Holder:2003eb}, as shown in
Figure~\ref{fig:reion_wrongparams}. Since the small scale $C_l$ scales
with the amplitude as $A_s e^{-2\tau}$, an incorrect optical depth
also means that the CMB constraint on the amplitude (and hence
e.g. $\sigma_8$) would also be incorrect. Modelling the reionization
history therefore has the joint aims of learning as much as we can
about how the universe reionized, at the same time as allowing other
cosmological parameters to be constrained reliably.

\subsection{Binning priors}
It is important to be aware of ones priors when using different
parameterizations. In particular, when binning some function,
introducing many new parameters with flat amplitude priors can
correspond to unintended priors on other parameters. For example,
consider the case when we somehow measure only the total optical depth
$\tau$. When $\tau$ is the only parameter we might get $\tau=\tau_0\pm
\sigma$, and for simplicity suppose the posterior is Gaussian. Now say
we bin the reionization history in $N$ bins that contribute equally to
the optical depth so that $\tau = \sum_n \tau_n$. Then if each bin
posterior is uncorrelated and Gaussian we have $\tau_n = \tau_0/N\pm
\sigma/\sqrt{N}$: we have measured each bin to have about the same
small contribution to the optical depth without having any relevant
data! This reflects the much larger parameter space volume where all
the bins have approximately the same contribution than when most of
the contribution is coming from just a few bins.

Looking at it the other way, consider $N$ bins with total optical
depth $\tau = \sum_n \tau_n$
where $0<\tau_n<\tau_n^{(1)}\equiv \tau_n(x_e=1)$, so we treat the
contributions to $\tau$ as though independent. If we have a flat prior
on $x_e$ in each bin with $0\le x_e\le1$, we have a flat prior on the
optical depth from each bin of $P(\tau_n)=1/\tau_n^{(1)}$ for
$0\le\tau_n\le\tau_n^{(1)}$ (and zero otherwise).
The prior on the total $\tau$ is then
\begin{eqnarray}
P(\tau) &=&  \int\ud\tau_0\dots \int \ud\tau_N\, \delta\left(\tau -
\sum_n \tau_n\right)\prod_{n=1}^N P(\tau_n)
\nonumber\\ &=& \int_{-\infty}^\infty \frac{\ud k}{2\pi} e^{i k \tau} \prod_{n=1}^N \left[  \int \ud \tau_n P(\tau_n) e^{-i k\tau_n}\right] \nonumber\\
&=&\int_{-\infty}^\infty \frac{\ud k}{2\pi} e^{i k \tau} \prod_{n=1}^N \left[ \frac{e^{-ik\tau_n^{(1)}}-1}{-ik\tau_n^{(1)}}\right] \nonumber \\
&\sim& \frac{1}{\sqrt{2\pi\sigma^2}}e^{-(\tau-\bar{\tau})^2/2\sigma^2}.
\end{eqnarray}
The exact integral gives a prior for $\tau$ that is piecewise
polynomial, peaking at $\bar{\tau}\equiv \sum_n \tau_n^{(1)}/2$ and
with variance $\sigma^2 \equiv \sum_n\tau_n^{(1)}{}^2/12$. This
approximates a Gaussian for large $N$. For bins with equal
$\tau^{(1)}$ the exact result is
\begin{equation}
P(\tau) = \sum_{n=0}^N \frac{(\tau/\tau^{(1)}-n)^N}{|\tau-n\tau^{(1)}|} \frac{(-1)^n N}{2 (N-n)! n!}.
\end{equation}
Assuming we are in the region where the posterior favours
$\tau<\bar{\tau}$, the binning prior will therefore favour higher
values of $\tau$ than using a flat prior on $\tau$. For $\tau$ close
to zero the prior goes like $\tau^{N-1}$: you are very unlikely to get
all the bins very close to zero. A good data constraint generally has
an exponential likelihood function, so the prior is only
logarithmically important compared to the log likelihood, however for
large numbers of bins the prior can have a significant effect.
In the absence of a strong constraint from the data, assuming a flat
prior on a large set of reionization bins will therefore favour
posterior constraints where each bin has a small contribution to the
optical depth and the total $\tau$ is higher than you would get from
no binning. It is important to use as much prior information as
possible to get relevant results: \emph{If you use a prior you don't
  believe, you shouldn't in general believe the posterior either}.
\begin{figure}
\begin{center}
\psfig{file=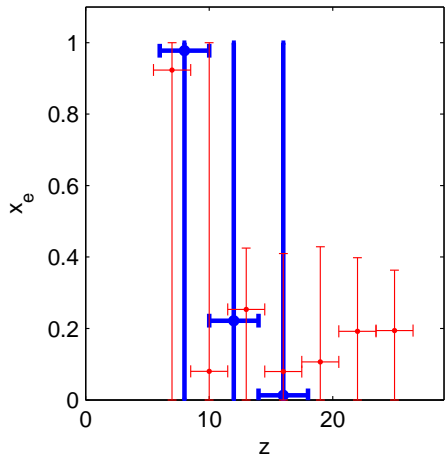,width=6cm}
\end{center}
\caption{Reionization fraction from WMAP 3-year data using seven bins
  with $\Delta z=3$ (thin lines) and three bins with $\Delta z=4$
  (thick lines, $x_e$ assumed zero at $z>18$). Vertical error bars
  show $68\%$ confidence regions. The results are largely prior
  driven, but bin correlations are $\alt 0.2$.}
 \label{fig:WMAP3}
\end{figure}

\subsection{Results}

Figure~\ref{fig:WMAP3} shows the constraint from current WMAP data
\citep{Hinshaw:2006ia,Page:2006hz} when three and seven reionization
bins are used. The WMAP polarization results are noise dominated, so
the constraint is very weak. As shown in the figure, the conclusions
one might come to depend on the prior: using more bins indicates that
the ionization fraction is constrained to be lower at intermediate
redshift than when using only three, if one considers the allowed
$1-\sigma$ regions. With only WMAP data the result is
largely prior driven. Since most people's prior does not correspond to
a random set of amplitudes in seven bins (e.g. most people probably
expect at most one local maximum), most people should not believe the
result. When better data is used the effect of an unbelievable
prior is much less important, but can still pull results in a
direction you don't really believe. An alternative parameterization
using a bin with a free $z_{\text{max}}$ is described in
\citet{Spergel:2006hy}, though this is not significantly constrained
by the data either (as expected ~\citep{Kaplinghat:2002vt}).

\begin{figure*}
\epsfig{file=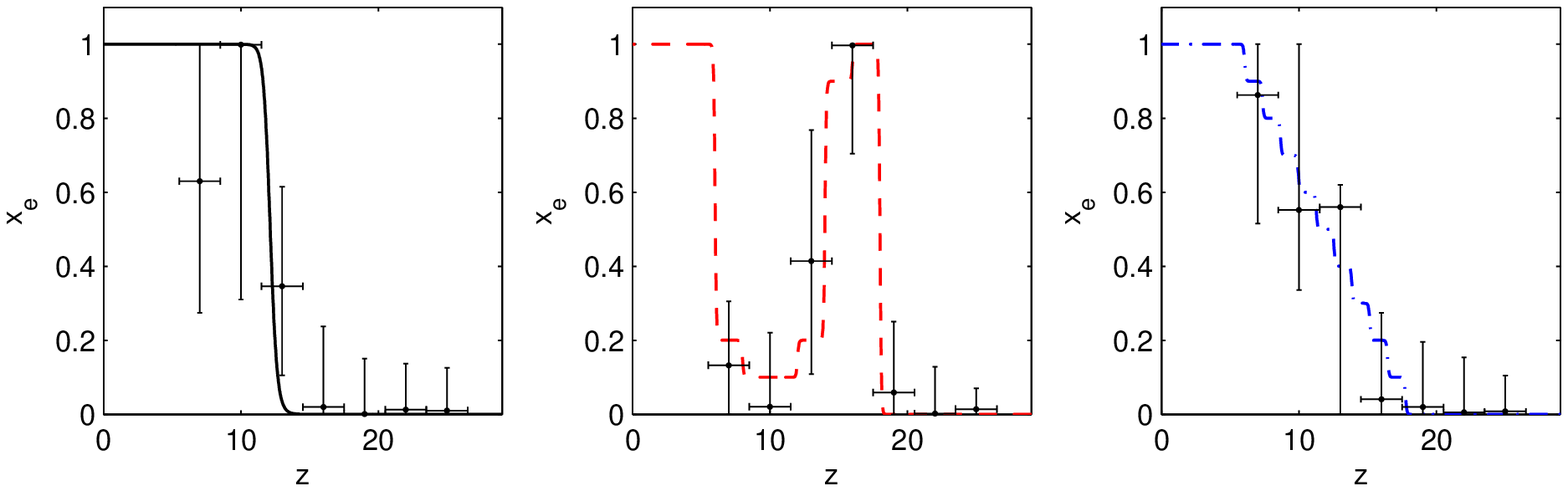,width=16cm}
\epsfig{file=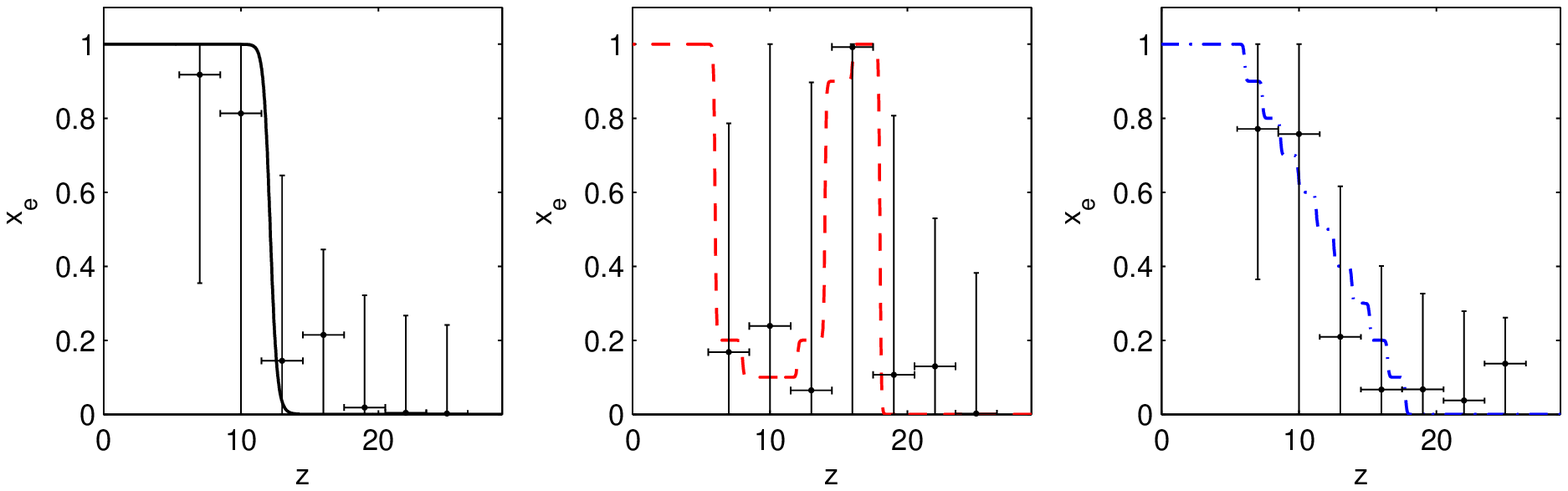,width=16cm}
\epsfig{file=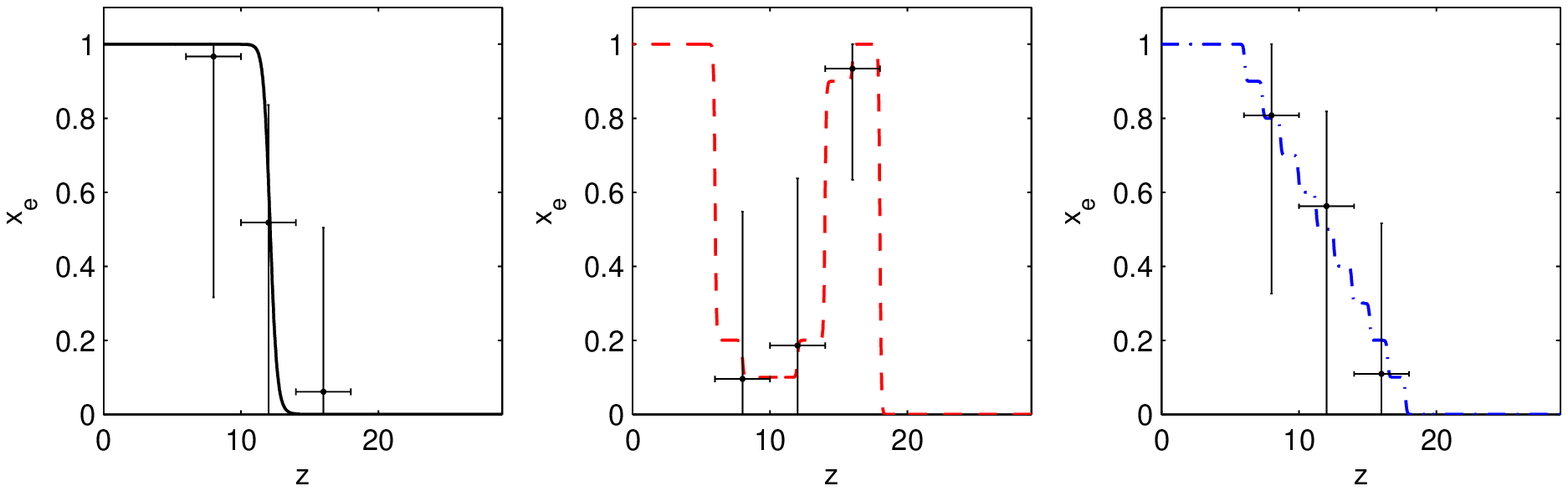,width=16cm}
\caption{Reconstructed reionization bins from perfect data (top) and
  \Planck\ (middle and bottom) for the three
fiducial models. Points show maximum sample likelihood positions, vertical error bars
are the marginalized $95\%$ confidence intervals, and the horizontal bars show the
width of the redshift bins ($\Delta z =3$ top two plots, $\Delta z =4$
  for bottom plot). The lines show the fiducial input model ionization
  history in each case. The bottom panel shows a \Planck\
  reconstruction for three bins with a prior that $x_e=0$ at $z>18$.
   Adjacent bins away from zero are correlated at the $0.3$--$0.7$ level.
}
\label{fig:recon}
\end{figure*}

In order to test how well CMB observations can constrain reionization
in principle we assume that all cosmological parameters, apart from
the bins in the ionization
fraction, are fixed. We then use the mean log likelihood from
noise-free full-sky data to forecast the reconstruction constraints,
using data from $l\le 100$ where the reionization signal is
important. We use $\zmin
= 7$, $\zmax = 25$ and $\Delta z = 3$. In the top panel of
Figure~\ref{fig:recon} we show the reconstructed ionization
histories for the three fiducial models. From these plots we see that
for perfect data we can distinguish the double reionization scenario
from the other two fiducial models at many sigma. Note however that
the error bars are correlated, and any detailed hypothesis test should
take this into account or work directly from the chains. Models can be
distinguished even if they appear consistent in a marginalized
error-bar plot: the regions in the full $n$-dimensional parameter space
may be different even if the projections are the same. A reionization
binning is a nice way to see what sharp redshift information is
available, but models that differ by broader features (for example the
sharp and dragged models) may be better modelled  using a more
targeted parameterization --- for example a couple of parameters
governing the slope of $x_e$ as it goes from $x_e=1$ at $z\sim 6$ to
zero at higher redshift.

We next include noise expected from the \Planck\
satellite and allow the other cosmological parameters to vary. The
result is shown in the middle panel of Figure \ref{fig:recon}. It is
hard to distinguish the three fiducial models at the
2-$\sigma$ level by comparing a number of marginalized reionization
redshift bin constraints. This is largely because there is a
degeneracy between the bins over the reionization peak: \Planck\
cannot resolve the redshift of the start of reionization accurately
enough to distinguish high $x_e$ followed by low $x_e$ from the case
where two bins contribute more equally. This is apparent in the (anti-)correlation of the error bars: when $x_e$ is significant there is a $\sim 0.3$ anti-correlation between adjacent points. If we include a prior that
$x_e=0$ at $z>18$, the constraint using $\Delta z=4$ bins looks much
better (bottom panel of Figure  \ref{fig:recon}), though the bins are
now fine-tuned to a priori knowledge about the expected form of the
history. Bin anti-correlations are still large at $\sim 0.5$--$0.7$. Our conclusions are not changed significantly if we instead
fix the cosmological parameters: \Planck\ constrain the other
parameters well, and on large scales the reconstruction is noise and
cosmic variance limited.

The constraint on $\tau$ when seven bins are used is consistent with
the fiducial value, whichever ionization history is used: see
Figure~\ref{fig:reion_wrongparams}.
Note that our analysis is assuming statistical isotropy; if the large scale
CMB is not isotropic as suggested by numerous analyses of the WMAP
data, it may be much harder to model or learn anything about the late
time ionization history. Alternatively it is possible that the signal
from a well understood ionization history could be used to learn more
about the structure of the large scale universe.

An alternative to our binned parameterization would be to use a
reduced set of well constrained Fisher PCA components identified by
\citet{Hu:2003gh}. However these have unphysical negative ionization
fractions, and the physical interpretation is clearer in easily
distinguishable models using our direct redshift binning. Also the PCA
components depend on the fiducial model (or an iterative scheme), so
a one-off binned parameter run is simpler to use in practice. We have
investigated the extraction of PCA components from our posterior bin
parameter chains, however the results we obtained appear not to be
very useful and rather different from the Fisher components of
\citet{Hu:2003gh}. This can be partly explained because infinitesimal
derivatives about a fiducial model generally give different covariance
estimates than samples from posteriors generated using physical priors
(e.g. that $0\le x_e\le 1$) accounting for the full non-Gaussian
posterior shape.

\section{Conclusions}

Modelling the ionization history is crucial to interpret high
precision observations of the CMB correctly. Uncertainties in the
details of recombination affect the small scale anisotropies, and an
incorrect model can lead to strongly biased parameter constraints. We
have made a first step towards modelling recombination uncertainties
using an ad hoc parameterization governing the recombination rate at
four different ionization fractions. This was good enough to recover
parameter estimates that are consistent, but at the cost of larger
error bars. There is clear motivation for future work to pin down the
recombination history in more detail theoretically, with
quantification of errors on any poorly measured important
parameters. In principle future work could do a full multi-atom
calculation for each cosmological model, including free parameters
with error bars for all rates that are not known very accurately. In
practice it is likely to be possible to devise an accurate small set
of effective equations that can be evolved quickly, along the lines of
the current \RECFAST\ code.
Using a free parameterization has the advantage of being able to look
for any surprises, for example energy injection from annihilating
particles, however if the standard model predictions are not well
nailed down it may be hard to distinguish subtle unexpected effects
from uncertainties in the expected model.

The reionization impacts the large scale CMB polarization
anisotropies. Our conclusions effectively agree with previous work,
though our direct binning approach is somewhat different. Ideal data
can do quite well at distinguishing distinct models, and even \Planck\
may be able to separate the most clear-cut cases.  We concentrate on
Planck, also there are ground based probes which could provide
useful information about reionization. In particular the Background
Imaging of Cosmic Extragalactic Polarization (BICEP) has in principle
the ability to measure features in the first trough in the E-mode
anisotropies, which at least to some degree could constrain the details
of the reionization history \citep{Keating:03}. However the error
bars remain quite large even with ideal data, and the CMB is
ultimately not going to be the best way to learn about the
reionization history observationally: 21cm emission is likely to be a
much better tracer. Allowing for different models in a CMB analysis is
however crucial to obtain unbiased constraints on the optical depth
(and hence the underlying perturbation variance) if the optical depth
is quite large. Of course if the optical depth is observed to be low,
so we know reionization must all have happened around $z\sim 6$, the
modelling for the history becomes less important as there is then
little room more complicated reionization scenarios. Alternatively a
consensus may arise that the reionization history is expected to be
monotonic (e.g. see \citet{Furlanetto:2004nt}), in which case modelling is less important and significantly
simpler.

\section{Acknowledgements}
We thank Douglas Scott, Wan Yan Wong, Steven Gratton, George
Efstathiou and Anthony Challinor for discussion and Xuelei Chen for
his modified version of \CAMB.
Most computations were performed on CITA's Mckenzie
cluster~\citep{Dubinski:2003vk} which was funded
  by the Canada Foundation for Innovation and the Ontario Innovation
  Trust. We acknowledge the use of the Legacy Archive for Microwave Background Data Analysis\footnote{\url{http://lambda.gsfc.nasa.gov/}} (LAMBDA). Support for LAMBDA is provided by the NASA Office of Space Science.  AL is supported by a PPARC Advanced Fellowship.

\providecommand{\aj}{Astron. J. }\providecommand{\apj}{Astrophys. J.
  }\providecommand{\apjl}{Astrophys. J.
  }\providecommand{\mnras}{MNRAS}\providecommand{\aap}{Astron. Astrophys.}

\label{lastpage}
\end{document}